\documentclass[preprint,showkeys]{revtex4}
\usepackage{graphicx}
\usepackage{dcolumn}
\usepackage{amsmath}
\usepackage{amssymb}
\usepackage{mathtools}
\usepackage{bm}
\newcommand{\be}{\begin{equation}}
\newcommand{\ee}{\end{equation}}
\newcommand{\ba}{\begin{eqnarray}}
\newcommand{\ea}{\end{eqnarray}}
\begin{document}
\title{On the accuracy of the melting curves \\
drawn from modelling a solid as an elastic medium}
\author{Santi Prestipino~\cite{aff1}}
\affiliation{Universit\`a degli Studi di Messina, Dipartimento di Fisica,
Contrada Papardo, I-98166 Messina, Italy}
\date{\today}
\begin{abstract}
An ongoing problem in the study of a classical many-body system is the
characterization of its equilibrium behaviour by theory or numerical
simulation. For purely repulsive particles, locating the melting line in
the pressure-temperature plane can be especially hard if the interparticle
potential has a softened core or contains some adjustable parameters.
A method is hereby presented that yields reliable melting-curve topologies
with negligible computational effort. It is obtained by combining the
Lindemann melting criterion with a description of the solid phase
as an elastic continuum. A number of examples are given in order to
illustrate the scope of the method and possible shortcomings.
For a two-body repulsion of Gaussian shape, the outcome of the present
approach compares favourably with the more accurate but also more
computationally demanding self-consistent harmonic approximation.
\end{abstract}
\maketitle

%
%
\section{Introduction}
There exists as yet no comprehensive theoretical treatment of the solid-liquid
phase transition that can rival in quality and accuracy the smooth-cutoff
formulation of the hierarchical reference theory of the liquid-vapour
transition~\cite{Parola}, which yields flat pressure vs. density isotherms
in the coexistence region as well as distinct binodal and spinodal curves.
Most theoretical and simulational strategies for detecting a point of
solid-liquid coexistence invariably pass through the prior determination
of the (Gibbs-) free energy for the two separate phases.
Theoretical approaches include classical density-functional theories and
thermodynamic perturbation theory~\cite{Hansen}. In Monte Carlo simulations,
the Frenkel-Ladd method and the Widom particle-insertion method may be
employed, in conjunction with thermodynamic integration, in order to obtain
accurate solid and liquid chemical potentials in the transition
region~\cite{Frenkel}.
On the far opposite side, lie a by now considerable number of empirical
one-phase melting and freezing criteria which allow a rough estimate of
the limit of stability for the given solid or liquid phase. Familiar
examples are the Lindemann melting rule and the Hansen-Verlet
freezing criterion, both relying on quantities that are computed numerically
(the mean square displacement in the solid and the structure factor in the
liquid).

I hereby consider a semi-empirical method for the melting transition which,
rather than being meant as a rule to provide a reliable estimate of the upper
stability threshold of a solid with prescribed symmetry, is actually aimed at
anticipating with little effort ({\it i.e.}, without resorting to
numerical simulation) at least the topology of the transition line, which
may be useful especially in those cases where multiple solid phases and/or
reentrant-fluid anomalies are expected~\cite{softpotentials}.
Assuming two-body forces between
the particles, the idea is to treat the solid system as an elastic medium
whose pressure-dependent moduli are determined at zero temperature from
the potential. The upper limit of thermodynamic stability of the solid is
then taken in accordance with the Lindemann rule~\cite{nota}.
Aside from the approximate character of the Lindemann criterion, the
main error in the estimate of the melting temperature $T_m(P)$ comes partly
from neglecting all anharmonicities in the particle dynamics and partly
from assuming the same elastic moduli at all temperatures. Both sources
of approximation are expected to extend solid stability well beyond the actual
threshold. In spite of this, the {\em shape} of $T_m(P)$ is reasonably well
reproduced by this criterion, as I shall demonstrate for a number of model
potentials (exceptions are anyway encountered -- see below).

After a reminder of elasticity theory in Section 2,
I introduce the novel criterion of melting in Section 3 along with a few
applications. In Section 4, I compare the indication of (what might be called)
the elastic criterion of melting for the repulsive Gaussian potential
with the outcomes of more refined approaches, based on the theory of the
harmonic crystal and on the self-consistent harmonic approximation.
Conclusions are postponed to Section 5.

%
%
\section{A brief account of linear elasticity}
\renewcommand{\theequation}{2.\arabic{equation}}
The information gathered here is standard reference material which is
preparatory to the theoretical analysis that will be outlined in the
next Section~\cite{elasticity}.

Consider the Bravais lattice $\{{\bf R}_n\}$ of a crystal with $N\gg 1$ atoms
(classical point particles) and let ${\bf x}_n={\bf R}_n+{\bf u}({\bf R}_n)$ be
the actual position of the $n$-th atom (to simplify the notation, I choose
${\bf R}_1=0$ in the following). In the simplest terms, the basic equations
of linear elasticity stem from evaluating the total potential energy $U$ of
the crystal in the approximation where ${\bf u}({\bf R})-{\bf u}({\bf R}')$
is expanded to linear order
in ${\bf R}-{\bf R}'$. For a homogeneous deformation, this amounts
to replacing $x_{n\alpha}$ with
\be
R_{n\alpha}+\sum_{\beta=1}^3u_{\alpha\beta}R_{n\beta}
\label{2-1}
\ee
for $\alpha=1,2,3$, $u_{\alpha\beta}$ being a constant. If the continuum
limit is taken, we can think of $u_{\alpha\beta}$ as just the constant
value of $\partial u_\alpha/\partial x_\beta$. Concurrently,
the deformation also modifies the crystal volume from $V_0$ to $V$:
\be
\frac{V}{V_0}=\det(\delta_{\alpha\beta}+u_{\alpha\beta})=
1+\sum_\alpha e_{\alpha\alpha}+\frac{1}{2}\sum_{\alpha,\beta}
(e_{\alpha\alpha}e_{\beta\beta}-
e_{\alpha\beta}^2)+\frac{1}{2}\sum_{\alpha,\beta}\omega_{\alpha\beta}^2\,,
\label{2-2}
\ee
where third-order terms in the strains were neglected. In Eq.\,(\ref{2-2}),
\be
e_{\alpha\beta}=\frac{1}{2}(u_{\alpha\beta}+u_{\beta\alpha})
\label{2-3}
\ee
is the {\em strain tensor} while
$\omega_{\alpha\beta}=(u_{\alpha\beta}-u_{\beta\alpha})/2$.
Our objective is to evaluate $U$ up to second order in the $u_{\alpha\beta}$.
Assuming a smooth spherically-symmetric pair potential $\phi(r)$ and
specializing the analysis to crystals of cubic symmetry, a
straightforward but tedious derivation yields:
\ba
u&\equiv&\frac{U}{N}=\frac{1}{2}\sum_{n=2}^N\phi(|{\bf x}_n|)=
u_0-P(v-v_0)+\frac{1}{2}v_0\lambda_1\left(e_{xx}^2+e_{yy}^2+e_{zz}^2\right)
\nonumber \\
&+&v_0\lambda_2\left(e_{xx}e_{yy}+e_{xx}e_{zz}+e_{yy}e_{zz}\right)+
2v_0\lambda_3\left(e_{xy}^2+e_{xz}^2+e_{yz}^2\right)\,,
\label{2-4}
\ea
with $v_0=V_0/N,u_0=(1/2)\sum_{n=2}^N\phi(|{\bf R}_n|)$, and
\be
P=-\frac{1}{6v_0}\sum_{n=2}^N|{\bf R}_n|\phi'(|{\bf R}_n|)\,.
\label{2-5}
\ee
Moreover, three {\em elastic constants} (or Lam\'e coefficients) appear in
Eq.\,(\ref{2-4}):
\ba
\lambda_1&=&-P+\frac{1}{2v_0}\sum_{n=2}^N\frac{X_n^4}{|{\bf R}_n|^4}
\left[|{\bf R}_n|^2\phi^{\prime\prime}(|{\bf R}_n|)-
|{\bf R}_n|\phi'(|{\bf R}_n|)\right]\,;
\nonumber \\
\lambda_2&=&P+\frac{1}{2v_0}\sum_{n=2}^N\frac{X_n^2Y_n^2}{|{\bf R}_n|^4}
\left[|{\bf R}_n|^2\phi^{\prime\prime}(|{\bf R}_n|)-
|{\bf R}_n|\phi'(|{\bf R}_n|)\right]\,;
\nonumber \\
\lambda_3&=&\lambda_2-2P\,,
\label{2-6}
\ea
where $X_n,Y_n$, and $Z_n$ are the Cartesian components of ${\bf R}_n$.
The three $\lambda$'s are the same quantities which are more commonly denoted
$c_{11},c_{12}$, and $c_{44}$, respectively. In Eq.\,(\ref{2-4}),
the term linear in the $u_{\alpha\beta}$ and actually proportional to the
trace of the strain tensor corresponds to the stress due to an applied
pressure $P$. Equation (\ref{2-5}) links the lattice parameter (or the
crystal volume $V_0$) with the pressure. The identification of $P$ with
the system pressure ensures consistency of Eq.\,(\ref{2-4})
with the thermodynamic definition of pressure.

A more general form of Eq.\,(\ref{2-4}), valid for any temperature $T$,
is the following:
\be
g=g_0+\frac{1}{2}v_0\sum_{\alpha,\beta,\gamma,\delta}
c_{\alpha\beta\gamma\delta}e_{\alpha\beta}e_{\gamma\delta}\,,
\label{2-7}
\ee
where $g$ is the Gibbs free energy per particle. Equation (\ref{2-7})
reduces to (\ref{2-4}) for $T=0$ and a crystal in the cubic system.
The maximum number of independent elastic constants
$c_{\alpha\beta\gamma\delta}$ is 21 (taking Voigt symmetry into account),
in fact they reduce to just three for crystals of cubic symmetry,
five for crystals of hexagonal symmetry, and so on. For instance,
for hexagonal solids Eq.\,(\ref{2-7}) takes the form
\ba
g&=&g_0+2v_0\lambda_1(e_{xx}+e_{yy})^2+v_0\lambda_2
\left[(e_{xx}-e_{yy})^2+4e_{xy}^2\right]
\nonumber \\
&+&\frac{1}{2}v_0\lambda_3e_{zz}^2+2v_0\lambda_4(e_{xx}+e_{yy})e_{zz}+
4v_0\lambda_5\left(e_{xz}^2+e_{yz}^2\right)\,,
\label{2-8}
\ea
with the following $T=0$ values of the Lam\'e coefficients:
\ba
\lambda_1&=&\frac{1}{12v_0}\sum_{n=2}^N\frac{X_n^4}{|{\bf R}_n|^4}
\left[|{\bf R}_n|^2\phi^{\prime\prime}(|{\bf R}_n|)-
|{\bf R}_n|\phi'(|{\bf R}_n|)\right]\,;
\nonumber \\
\lambda_2&=&\lambda_1-\frac{P}{2}\,;
\nonumber \\
\lambda_3&=&-P+\frac{1}{2v_0}\sum_{n=2}^N\frac{Z_n^4}{|{\bf R}_n|^4}
\left[|{\bf R}_n|^2\phi^{\prime\prime}(|{\bf R}_n|)-
|{\bf R}_n|\phi'(|{\bf R}_n|)\right]\,;
\nonumber \\
\lambda_4&=&\frac{P}{2}+\frac{1}{4v_0}\sum_{n=2}^N
\frac{X_n^2Z_n^2}{|{\bf R}_n|^4}
\left[|{\bf R}_n|^2\phi^{\prime\prime}(|{\bf R}_n|)-
|{\bf R}_n|\phi'(|{\bf R}_n|)\right]\,;
\nonumber \\
\lambda_5&=&\lambda_4-P\,.
\label{2-9}
\ea
For tetragonal crystals, one similarly finds
\ba
g&=&g_0+\frac{v_0}{2}\left[\lambda_1(e_{xx}^2+e_{yy}^2)+
2\lambda_2e_{xx}e_{yy}+4\lambda_3e_{xy}^2\right.
\nonumber \\
&+&\left.\lambda_4e_{zz}^2+2\lambda_5(e_{xx}+e_{yy})e_{zz}+
4\lambda_6\left(e_{xz}^2+e_{yz}^2\right)\right]\,,
\label{2-10}
\ea
with zero-temperature Lam\'e coefficients given by
\ba
\lambda_1&=&-P+\frac{1}{2v_0}\sum_{n=2}^N\frac{X_n^4}{|{\bf R}_n|^4}
\left[|{\bf R}_n|^2\phi^{\prime\prime}(|{\bf R}_n|)-
|{\bf R}_n|\phi'(|{\bf R}_n|)\right]\,;
\nonumber \\
\lambda_2&=&P+\frac{1}{2v_0}\sum_{n=2}^N\frac{X_n^2Y_n^2}{|{\bf R}_n|^4}
\left[|{\bf R}_n|^2\phi^{\prime\prime}(|{\bf R}_n|)-
|{\bf R}_n|\phi'(|{\bf R}_n|)\right]\,;
\nonumber \\
\lambda_3&=&\lambda_2-2P\,;
\nonumber \\
\lambda_4&=&-P+\frac{1}{2v_0}\sum_{n=2}^N\frac{Z_n^4}{|{\bf R}_n|^4}
\left[|{\bf R}_n|^2\phi^{\prime\prime}(|{\bf R}_n|)-
|{\bf R}_n|\phi'(|{\bf R}_n|)\right]\,;
\nonumber \\
\lambda_5&=&P+\frac{1}{2v_0}\sum_{n=2}^N\frac{X_n^2Z_n^2}{|{\bf R}_n|^4}
\left[|{\bf R}_n|^2\phi^{\prime\prime}(|{\bf R}_n|)-
|{\bf R}_n|\phi'(|{\bf R}_n|)\right]\,;
\nonumber \\
\lambda_6&=&\lambda_5-2P\,.
\label{2-11}
\ea

At $T=0$, the expansion of the Helmholtz free energy $F=Nf$ in powers of
the strain-tensor components is the same as for $U$. For non-zero
temperatures, the respective $c_{\alpha\beta\gamma\delta}$ are instead
different (one thus distinguishes isothermal and adiabatic elastic constants).
For any $T$, the Helmholtz free energy of a solid under arbitrary
initial stress can otherwise be expanded to second order in the components
of the displacement gradients $u_{\alpha\beta}$,
\be
\frac{f-f_0}{v_0}=\sum_{\alpha,\beta}S_{\alpha\beta}u_{\alpha\beta}+
\frac{1}{2}\sum_{\alpha,\beta,\gamma,\delta}S_{\alpha\beta\gamma\delta}
u_{\alpha\beta}u_{\gamma\delta}\,;
\label{2-12}
\ee
alternatively, $f$ can be written as a truncated power series of the
Lagrangian strain parameters,
\be
\eta_{\alpha\beta}=\frac{1}{2}\left(u_{\alpha\beta}+u_{\beta\alpha}+
\sum_\gamma u_{\gamma\alpha}u_{\gamma\beta}\right)\,,
\label{2-13}
\ee
with yet different coefficients in the linear and quadratic terms:
\be
\frac{f-f_0}{v_0}=\sum_{\alpha,\beta}C_{\alpha\beta}\eta_{\alpha\beta}+
\frac{1}{2}\sum_{\alpha,\beta,\gamma,\delta}C_{\alpha\beta\gamma\delta}
\eta_{\alpha\beta}\eta_{\gamma\delta}\,.
\label{2-14}
\ee
It is then a simple exercise to show that $S_{\alpha\beta}=C_{\alpha\beta}$
and $S_{\alpha\beta\gamma\delta}=C_{\alpha\beta\gamma\delta}+
C_{\beta\delta}\delta_{\alpha\gamma}$. Moreover, for
$C_{\alpha\beta}=-P\delta_{\alpha\beta}$, one finds that
\be
c_{\alpha\beta\gamma\delta}=C_{\alpha\beta\gamma\delta}+
P(\delta_{\alpha\beta}\delta_{\gamma\delta}-
\delta_{\alpha\gamma}\delta_{\beta\delta}-
\delta_{\alpha\delta}\delta_{\beta\gamma})\,.
\label{2-15}
\ee
Equation (\ref{2-15}) is useful for computing the $\lambda$'s at $T>0$
through numerical simulation
since specific virial-like formulae exist for the $C$'s~\cite{Farago}.

An important issue is mechanical stability of a crystal phase, which is a
prerequisite for its thermodynamic stability: an applied strain may
destabilize the crystal, which in this case is really
stable only at zero temperature. The elastic constants in Eq.\,(\ref{2-7})
must obey so-called stability conditions in order for the unstrained crystal
to resist any infinitesimal deformation,
{\it i.e.}, in order for the crystal lattice $\{{\bf R}_n\}$ to provide a
minimum (not just an extremum) for $g$. Depending on the interparticle
potential and on the pressure value, the crystal may or may not be
mechanically stable, meaning that it does typically exist as a stable
structure for $T > 0$ only within one or more definite pressure ranges.

Using Voigt symmetry, the
elastic constants of a cubic crystal can be arranged in the $6\times 6$ matrix
\be
\left(
\begin{array}{ccc|ccc}
\lambda_1 & \lambda_2 & \lambda_2 & 0 & 0 & 0 \\ 
\lambda_2 & \lambda_1 & \lambda_2 & 0 & 0 & 0 \\
\lambda_2 & \lambda_2 & \lambda_1 & 0 & 0 & 0 \\
\hline
0 & 0 & 0 & \lambda_3 & 0 & 0 \\
0 & 0 & 0 & 0 & \lambda_3 & 0 \\
0 & 0 & 0 & 0 & 0 & \lambda_3
\end{array}
\right)
\label{2-16}
\ee
and Eq.\,(\ref{2-7}) becomes a quadratic form,
$g=g_0+(v_0/2)\sum_{a,b=1}^6c_{ab}e_ae_b$
with $e_1=e_{xx},e_2=e_{yy},e_3=e_{zz},e_4=2e_{yz},e_5=2e_{xz},e_6=2e_{xy}$.
The eigenvalues (with multiplicities) of (\ref{2-16}) are $\lambda_3$ (3),
$\lambda_1-\lambda_2$ (2), and $\lambda_1+2\lambda_2$ (1), leading to three
stability conditions:
\be
\lambda_1+2\lambda_2\ge 0\,;\,\,\lambda_3\ge 0\,;\,\,
\lambda_1-\lambda_2\ge 0\,.
\label{2-17}
\ee
The first two conditions amount to requiring the existence of the bulk and
the shear modulus, respectively. The last inequality prescribes rigidity of
the cubic solid against tetragonal shear.
For a hexagonal crystal, a similar analysis yields four conditions,
\be
\lambda_2\ge 0\,;\,\,\lambda_5\ge 0\,;\,\,8\lambda_1+\lambda_3\ge 0\,;\,\,
\lambda_1\lambda_3-\lambda_4^2\ge 0\,,
\label{2-18}
\ee
becoming five for tetragonal crystals:
\be
\lambda_3\ge 0\,;\,\,\lambda_6\ge 0\,;\,\,\lambda_1-\lambda_2\ge 0\,;\,\,
\lambda_1+\lambda_2+\lambda_4\ge 0\,;\,\,
\lambda_4(\lambda_1+\lambda_2)-2\lambda_5^2\ge 0\,.
\label{2-19}
\ee

Tightly related to the subject of solid elasticity is the general harmonic
theory of lattice dynamics. Consider a finite crystal with externally applied
classical forces, and let the forces be restricted to the surface region so
as to represent stresses applied to the crystal. Since the total force on
each atom must vanish when the atoms are located at the equilibrium positions
$\{{\bf R}_n\}$, the total energy at $T = 0$ can be approximately written as
\be
U=U_0+\frac{1}{2}\sum_{{\bf R},{\bf R}'}\sum_{\alpha,\beta}
\Phi_{\alpha\beta}({\bf R}-{\bf R}')u_\alpha({\bf R})u_\beta({\bf R}')
\label{2-20}
\ee
with $U_0=U({\bf R}_1,\ldots ,{\bf R}_N)$, all anharmonicities being neglected.
The $\Phi$ coefficients in (\ref{2-20}) are second-order derivatives,
\be
\Phi_{\alpha\beta}({\bf R}-{\bf R}')=\left(\frac{\partial^2U}
{\partial u_\alpha({\bf R})\partial u_\beta({\bf R}')}\right)_0\,,
\label{2-21}
\ee
and, for a Bravais crystal, they are invariant under the exchange
$\alpha\leftrightarrow\beta$ because of the lattice inversion symmetry.
Invariance of the energy value following a rigid translation of the crystal
further leads to $\sum_{\bf R}\Phi_{\alpha\beta}({\bf R})=0$ for any
$\alpha$ and $\beta$.
 
The equations of motion for the potential energy (\ref{2-20}) read
\be
m\ddot{u}_\alpha({\bf R})=-\sum_{{\bf R}',\beta}
\Phi_{\alpha\beta}({\bf R}-{\bf R}')u_\beta({\bf R}')\,,
\label{2-22}
\ee
where $m$ is the particle mass, and are solved in terms of plane waves
(the {\em normal modes} of vibration),
\be
\epsilon_\alpha({\bf q})e^{i\left[{\bf q}\cdot{\bf R}-
\omega({\bf q})t\right]}\,\,\,\,(\alpha=1,2,3)\,.
\label{2-23}
\ee
The $N$ values of ${\bf q}$ lie within the first Brillouin zone (1BZ) of the
lattice and are so chosen as
to allow for the periodic repetition of the lattice outside its boundaries.
Upon introducing the (real symmetric) {\em dynamical matrix}
\be
B_{\alpha\beta}({\bf q})=\sum_{\bf R}
\Phi_{\alpha\beta}({\bf R})e^{i{\bf q}\cdot{\bf R}}=
-\sum_{\bf R}\Phi_{\alpha\beta}({\bf R})
\left[1-\cos({\bf q}\cdot{\bf R})\right]\,,
\label{2-24}
\ee
the normal-mode amplitudes are found to obey the linear set of equations
\be
m\omega^2({\bf q})\epsilon_\alpha({\bf q})=
\sum_\beta B_{\alpha\beta}({\bf q})\epsilon_\beta({\bf q})\,.
\label{2-25}
\ee
For any ${\bf q}$, the three eigenvalues of $B_{\alpha\beta}({\bf q})$,
namely $m\omega_s^2({\bf q})$
($s=1,2,3$), are real and we can always choose orthonormal eigenvectors,
$\sum_\alpha\epsilon_{s\alpha}({\bf q})\epsilon_{s'\alpha}({\bf q})=
\delta_{ss'}$.
The explicit form of the dynamical-matrix components is
$B_{\alpha\alpha}=\tau_{\alpha\alpha}-\tau_1$
and $B_{\alpha\beta}=\tau_{\alpha\beta}$ ($\alpha\neq\beta$), where
\ba
\tau_1({\bf q})&=&-\sum_{n\neq 1}\frac{\phi'(|{\bf R}_n|)}{|{\bf R}_n|}
\left[1-\cos({\bf q}\cdot{\bf R}_n)\right]\,;
\nonumber \\
\tau_{\alpha\beta}({\bf q})&=&\sum_{n\neq 1}
\frac{X_\alpha X_\beta}{|{\bf R}_n|^4}
\left[|{\bf R}_n|^2\phi''(|{\bf R}_n|)-
|{\bf R}_n|\phi'(|{\bf R}_n|)\right]
\left[1-\cos({\bf q}\cdot{\bf R}_n)\right]\,.
\label{2-26}
\ea

A crystal dynamics is also associated with the approximation set by linear
elasticity. It is drawn from the Lagrangian density (cf. Eq.\,(\ref{2-12}))
\be
{\cal L}=\frac{1}{2}\rho\,\dot{u}^2({\bf x})-\sum_{\alpha,\beta}
S_{\alpha\beta}u_{\alpha\beta}-\frac{1}{2}\sum_{\alpha,\beta,\gamma,\delta}
S_{\alpha\beta\gamma\delta}u_{\alpha\beta}u_{\gamma\delta}\,,
\label{2-27}
\ee
where $\rho$ is the mass density. From Eq.\,(\ref{2-27}) one derives the
equations of motion
\be
\rho\ddot{u}_\alpha({\bf x})=\sum_{\beta,\gamma,\delta}
c_{\alpha\beta\gamma\delta}
\frac{\partial^2u_\gamma}{\partial x_\beta\partial x_\delta}\,,
\label{2-28}
\ee
whose solutions of are still plane waves with frequencies given by the
secular equation
\be
\det\left\{\sum_{\beta\delta}c_{\alpha\beta\gamma\delta}q_\beta q_\delta-
\rho\,\omega^2({\bf q})\delta_{\alpha\gamma}\right\}=0\,.
\label{2-29}
\ee
In particular, one observes that the elastic waves are dispersionless,
{\it i.e.}, $\omega^2\propto q^2$.
For cubic crystals, the explicit form of Eq.\,(\ref{2-29}) is:
\be
\left|
\begin{array}{ccc}
c_{11}q_x^2+c_{44}(q_y^2+q_z^2)-\rho\omega^2 & (c_{12}+c_{44})q_xq_y &
(c_{12}+c_{44})q_xq_z \\ 
(c_{12}+c_{44})q_xq_y & c_{11}q_y^2+c_{44}(q_x^2+q_z^2)-\rho\omega^2 &
(c_{12}+c_{44})q_yq_z \\
(c_{12}+c_{44})q_xq_z & (c_{12}+c_{44})q_yq_z &
c_{11}q_z^2+c_{44}(q_x^2+q_y^2)-\rho\omega^2
\end{array}
\right|=0\,.
\label{2-30}
\ee

%
%
\section{A new melting criterion}
\setcounter{equation}{0}
\renewcommand{\theequation}{3.\arabic{equation}}
In this Section, a Gaussian field theory is formulated in order to describe
the thermal properties of an elastic solid in the simplest possible terms.
The aim is to obtain an approximate value for the mean square displacement
(MSD) of crystal atoms that can be used to estimate the melting temperature
of the crystal through the Lindemann criterion.

Consider for concreteness a crystal of cubic symmetry with $N=N_1N_2N_3$
atoms. Rather than assuming a homogeneous strain, I allow for a spatial
dependence of atomic displacements and take the continuum limit. Then,
the enthalpy $H$ of the crystal at $T=0$ becomes (cf. Eq.\,(\ref{2-4})):
\ba
H=H_0&+&\frac{1}{2}\int_{V_0}{\rm d}^3r\left\{\lambda_1\left[
\left(\frac{\partial u_x}{\partial x}\right)^2+
\left(\frac{\partial u_y}{\partial y}\right)^2+
\left(\frac{\partial u_z}{\partial z}\right)^2\right]\right.
\nonumber \\
&+&2\lambda_2\left(
\frac{\partial u_x}{\partial x}\frac{\partial u_y}{\partial y}+
\frac{\partial u_x}{\partial x}\frac{\partial u_z}{\partial z}+
\frac{\partial u_y}{\partial y}\frac{\partial u_z}{\partial z}\right)
\nonumber \\
&+&\left.\lambda_3\left[
\left(\frac{\partial u_x}{\partial y}+\frac{\partial u_y}{\partial x}\right)^2+
\left(\frac{\partial u_x}{\partial z}+\frac{\partial u_z}{\partial x}\right)^2+
\left(\frac{\partial u_y}{\partial z}+\frac{\partial u_z}{\partial y}\right)^2
\right]\right\}
\label{3-1}
\ea
with the $\lambda$'s given by Eq.\,(\ref{2-6}). Upon implementing periodic
boundary conditions, the displacement vector is expanded in a series of
plane waves:
\be
u_\alpha({\bf r})=\sum_{\bf q}\widetilde{u}_\alpha({\bf q})
e^{i{\bf q}\cdot{\bf r}}\,\,\,\,{\rm (conversely,}\,\,
\widetilde{u}_\alpha({\bf q})=\frac{1}{V_0}\int_{V_0}{\rm d}^3r\,
u_\alpha({\bf r})e^{-i{\bf q}\cdot{\bf r}}{\rm )}\,,
\label{3-2}
\ee
where, in terms of reciprocal-lattice primitive vectors, the wave vector
${\bf q}=\sum_\alpha q_\alpha{\bf b}_\alpha$ with
$q_\alpha=m_\alpha/N_\alpha$ and $m_\alpha=-N_\alpha/2+1,\ldots,N_\alpha/2$
($\alpha=1,2,3$). Substitution of (\ref{3-2}) into (\ref{3-1}) leads
eventually to
\be
H=H_0+\frac{1}{2}V_0\sum_{\bf q}\sum_{\alpha,\beta}A_{\alpha\beta}({\bf q})
\widetilde{u}_\alpha({\bf q})\widetilde{u}_\beta^*({\bf q})
\label{3-3}
\ee
with
\be
A_{\alpha\beta}({\bf q})=\left[\lambda_3q^2+
(\lambda_1-\lambda_2-2\lambda_3)q_{\alpha}^2\right]\delta_{\alpha\beta}+
(\lambda_2+\lambda_3)q_{\alpha}q_{\beta}\,.
\label{3-4}
\ee
Next, I try to represent the thermal disordering of the crystal through a
field theory where the basic variables are the $u_\alpha({\bf r})$'s and the
statistical weight of field configurations is $\exp(-\beta H)$. This choice
is tantamount to the assumption of $T$-independent elastic
constants, whose values are fixed at their ($P$-dependent) $T=0$ values.

To compute the MSD, the following average is to be evaluated first:
\be
\left<\widetilde{u}_\alpha({\bf q})\widetilde{u}_\beta^*({\bf q})\right>=
\frac{\int{\cal D}\widetilde{u}{\cal D}\widetilde{u}^*\,
\widetilde{u}_\alpha({\bf q})\widetilde{u}_\beta^*({\bf q})
\exp\{-\beta V_0\sum_{{\bf q}>0}\sum_{\gamma,\delta}A_{\gamma\delta}({\bf q})
\widetilde{u}_\gamma({\bf q})\widetilde{u}_\delta^*({\bf q})\}}
{\int{\cal D}\widetilde{u}{\cal D}\widetilde{u}^*\,
\exp\{-\beta V_0\sum_{{\bf q}>0}\sum_{\gamma,\delta}A_{\gamma\delta}({\bf q})
\widetilde{u}_\gamma({\bf q})\widetilde{u}_\delta^*({\bf q})\}}\,,
\label{3-5}
\ee
where in both integrals the ${\bf q}$'s are restricted to half space
(symbolically, ${\bf q}>0$) in order that
$\{{\rm Re}\,\widetilde{u}_\alpha({\bf q}),
{\rm Im}\,\widetilde{u}_\alpha({\bf q})\}$
can be treated as independent integration variables -- namely,
${\cal D}\widetilde{u}{\cal D}\widetilde{u}^*=
\prod_{{\bf q}>0}\prod_\alpha
{\rm d}\left({\rm Re}\,\widetilde{u}_\alpha({\bf q})\right)\,
{\rm d}\left({\rm Im}\,\widetilde{u}_\alpha({\bf q})\right)$.
Using properties of
complex-valued Gaussian integrals, one obtains
\be
\left<\widetilde{u}_\alpha({\bf q})\widetilde{u}_\beta^*({\bf q})\right>=
\frac{k_BT}{V_0}\left(A^{-1}\right)_{\alpha\beta}({\bf q})\,.
\label{3-6}
\ee
Since the inverse of a symmetric matrix is also symmetric, the previous
result actually applies for any ${\bf q}$. Hence, the MSD reads
\be
\left<\frac{1}{V_0}\int_{V_0}{\rm d}^3r\,u^2({\bf r})\right>=
\sum_{\bf q}\sum_\alpha\left<\left|\widetilde{u}_\alpha({\bf q})
\right|^2\right>=\frac{k_BT}{V_0}\sum_{\bf q}{\rm Tr}A^{-1}({\bf q})\,.
\label{3-7}
\ee
In the thermodynamic limit, the residual sum transforms into an integral
over the 1BZ,
which is more easily computed by replacing the zone with a (Debye)
sphere of equal volume (the error committed is small), with the result:
\be
\left<\frac{1}{V_0}\int_{V_0}{\rm d}^3r\,u^2({\bf r})\right>=
\frac{k_BT}{\pi^3}q_D\int_0^{\pi/2}{\rm d}\phi\int_0^{\pi/2}{\rm d}\theta\,
\sin\theta\,\frac{f_1(\theta,\phi)}{f_2(\theta,\phi)}\,,
\label{3-8}
\ee
where $q_D=(6\pi^2\rho)^{1/3}$ and
\ba
f_1(\theta,\phi)&=&\lambda_3(\lambda_3+2\lambda_1)+(\lambda_1+\lambda_2)
(\lambda_1-\lambda_2-2\lambda_3)
(\sin^4\theta\sin^2\phi\cos^2\phi+\sin^2\theta\cos^2\theta)\,;
\nonumber \\
f_2(\theta,\phi)&=&\lambda_1\lambda_3^2+\lambda_3(\lambda_1+\lambda_2)
(\lambda_1-\lambda_2-2\lambda_3)(\sin^4\theta\sin^2\phi\cos^2\phi+
\sin^2\theta\cos^2\theta)
\nonumber \\
&+&(\lambda_1-\lambda_2-2\lambda_3)^2(\lambda_1+2\lambda_2+\lambda_3)
\sin^4\theta\cos^2\theta\sin^2\phi\cos^2\phi\,.
\label{3-9}
\ea
The parallel treatment for a harmonic crystal moves from
\be
U=U_0+\frac{N}{2}\sum_{\bf q}\sum_{\alpha,\beta}B_{\alpha\beta}({\bf q})
\widetilde{u}_\alpha({\bf q})\widetilde{u}_\beta^*({\bf q})\,,
\label{3-10}
\ee
and eventually leads, through the same series of steps as before,
to the following expression for the MSD,
\be
\left<\frac{1}{N}\sum_{\bf R}u^2({\bf R})\right>=\frac{k_BT}{(2\pi)^3}v_0
\int_0^{q_D}{\rm d}q\,q^2\int{\rm d}^2\Omega\,{\rm Tr}B^{-1}({\bf q})\,,
\label{3-11}
\ee
which is more numerically demanding than (\ref{3-8}) because of the additional
$q$ integration present in (\ref{3-11}).

We see from Eqs.\,(\ref{3-8}) and (\ref{3-11}) that the MSD increases linearly
with $T$. According to the Lindemann criterion, the crystal melts when the
MSD reaches a fraction $L_m\approx 0.1$ of the nearest-neighbour distance
$a_{NN}$, from which the estimate of $T_m(P)$ follows directly.
For face-centred cubic (fcc), hexagonal close-packed (hcp), and body-centred
cubic (bcc) crystals, the specific $L_m$ values are $0.15,0.10$,
and $0.18$, respectively~\cite{Saija,Cho}, while
no systematic study of the typical values of the Lindemann ratio for other
crystals has ever been undertaken, at least to my knowledge (hence, I assume
$L_m=0.1$ indifferently for all such phases). If any of the stability
conditions is violated for a crystal under pressure $P_0$,
then I take a zero melting temperature for the given solid at $P=P_0$.

Universality of $L_m$ along the fluid-solid coexistence line is well
established for fcc, bcc, and hcp crystals. For other types of crystals no
such information is available and this makes the $T_m$ estimated through what
I shall call {\em the elastic criterion of melting} less reliable for these
crystals. In general,
the elastic constants get smaller and smaller on increasing temperature until
they abruptly vanish on crossing the melting line. Hence, assuming the elastic
constants to be independent of $T$ is a major simplification that leads to
systematically underestimating the MSD; moreover, also the neglect
of anharmonic terms in the potential would likely contribute to enhancing the
stability of the solid, with the effect that the $T_m(P)$ computed with the
elastic criterion of melting will be larger than the actual value.
One may reasonably
expect that the extent to which the melting temperature is overestimated
is roughly the same for all pressures so that at least the shape of $T_m(P)$
is got correctly.

A first application of the elastic criterion is to the melting of the
Lennard-Jones fluid, which is known to crystallize into a hcp solid
(unless the pressure is huge -- larger than 800 in reduced $\epsilon/\sigma^3$
units). For reduced pressures smaller than 20, the computed $T_m$ is a
concave function of $P$, as expected~\cite{Mastny}.
For $P=1$ and $P=10$, the criterion predicts a melting temperature of 1.18
and 1.75, respectively, whereas the ``exact'' values from Ref.\,\cite{Mastny}
are 0.78 and 1.40.

A more challenging test of the elastic criterion is offered by a recent
simulation study~\cite{Prestipino} of a system of particles repelling
each other through the Yoshida-Kamakura (YK) potential,
\be
\phi_{\rm YK}(r)=\epsilon\exp\left[a\left(1-\frac{r}{\sigma}\right)-
6\left(1-\frac{r}{\sigma}\right)^2\ln\frac{r}{\sigma}\right]
\label{3-12}
\ee
with $a=3.3$. For reduced pressures smaller than 3, the phase diagram of the
model is plotted in figure 3 of Ref.\,~\cite{Prestipino}.
The same phase diagram but computed through the present melting criterion
(with an
enormous saving of time compared to simulation) is reported in Fig.\,1. Here
are shown the melting lines for a number of solid phases chosen among those
stable at zero temperature. For each crystal, the melting curve is a
single line or it consists of a number of disjoint pieces, one for each
range of pressure/density where the stability conditions are met. It is
worth stressing that the pressure range of mechanical stability of
a phase is usually wider than the range of {\em thermodynamic} stability
at $T=0$, which is where the enthalpy of the phase is smaller than that of
any other crystal phase. Hence, the stability boundaries dictated by the
elastic criterion do not generally coincide with the actual thermodynamic
thresholds.

On approaching a stability boundary, the MSD of Eq.\,(\ref{3-8}) blows up and
the melting temperature drops continuously to zero. The line of fluid-solid
coexistence would correspond to the upper envelope of the melting curves
for the various solids. It is clear from Fig.\,1 that the gross features of
the phase diagram of the YK fluid are well reproduced by the elastic
criterion, the main error being in the regular overestimation of the melting
temperature. The greater stability of the $\beta$-Sn phase over the
simple hexagonal (sh) solid
in the pressure range between roughly 3 and 7 might be just accidental,
related to the choice of the same $L_m$ for both. The harmonic approximation
works quantitatively better (since at variance with linear elasticity no
large-wavelength limit is implied) but it takes a much longer computer time
to calculate the MSD.

\begin{figure}
\includegraphics[width=8cm]{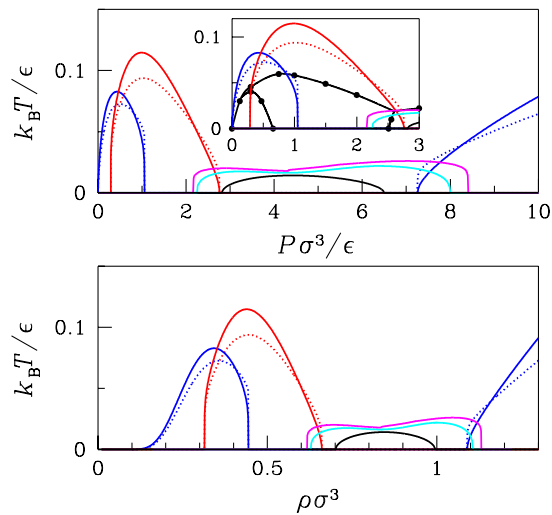}
\caption{(Color online). Schematic phase diagram of the YK potential with
$a=3.3$ as drawn from the elastic criterion of melting. The melting lines
of various solid phases are shown: fcc (blue), bcc (red), simple cubic
(sc, black), sh (cyan), and $\beta$-Sn (magenta). The dotted lines are the
melting curves for the fcc and bcc crystals as derived from the harmonic
approximation, see Eq.\,(\ref{3-11}). In the inset (top panel), a
comparison is made with the exact coexistence boundaries of the model
(black dots and thick solid lines)~\cite{Prestipino}. From low to high
pressure, the stable phases up to $P=3$ are fcc, bcc, and $\beta$-Sn.}
\label{fig1}
\end{figure}

To better appreciate the quality of the elastic criterion of melting,
it is worth considering what would be the phase diagram of the YK
potential with $a=3.3$ according to a theory of fluid-solid coexistence
based on the use of the cell-theory approximation for the solid and the
Mansoori-Canfield theory for the fluid (see the details in the Appendix).
We see from Fig.\,2 that this theory predicts a direct transition from bcc
to sh at high temperature, a possibility which was not actually considered
in the simulation; however, the melting temperature of the YK fluid is
overestimated by the theory to roughly the same extent ($\approx 100$\%)
as it is by the elastic criterion, a fact that alone casts some shadows
on the reliability of the theoretical phase diagram.

\begin{figure}
\includegraphics[width=8cm]{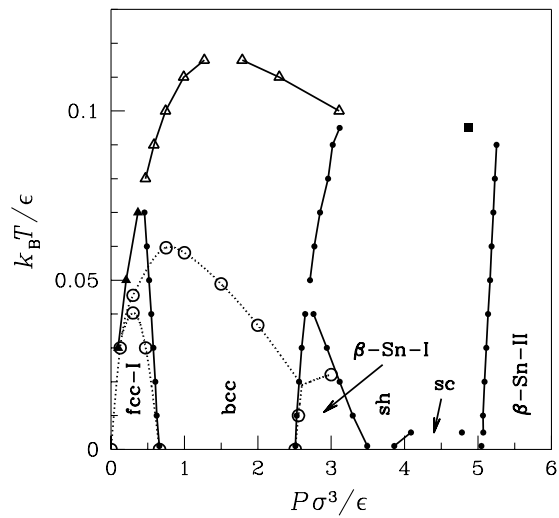}
\caption{Low-pressure phase diagram of the YK potential with $a=3.3$
according to the theory detailed in the Appendix. Solid-solid coexistence
points are depicted as small dots, whereas triangles and the full square
are solid-fluid coexistence points. When there are more than one crystal
phase of a given type, a Roman numeral distinguishes between them
(e.g. $\beta$-Sn-I and $\beta$-Sn-II; the second fcc phase is stable
for pressures out of the range shown). The dotted lines through the open
dots are the coexistence loci of the model from Ref.\,\cite{Prestipino}.}
\label{fig2}
\end{figure}

It is instructive to look at the shape of some representative phonon branches
of the bcc crystal of YK particles for $\rho=0.6607$ ($P\simeq 2.76$),
{\it i.e.}, where the bcc solid is about to become unstable at zero
temperature owing to the fact that $c_{44}$ is almost zero and actually
negative for larger pressures. This instability is caused by phonon softening
at the $\Gamma$ point: along the path from $\Gamma$ to N, one of the
phonon branches satisfies
$m\omega^2({\bf q})\simeq(c_{44}/\rho)(q_x^2+q_y^2)$ for $q\rightarrow 0$
(see Fig.\,3).

\begin{figure}
\includegraphics[width=8cm]{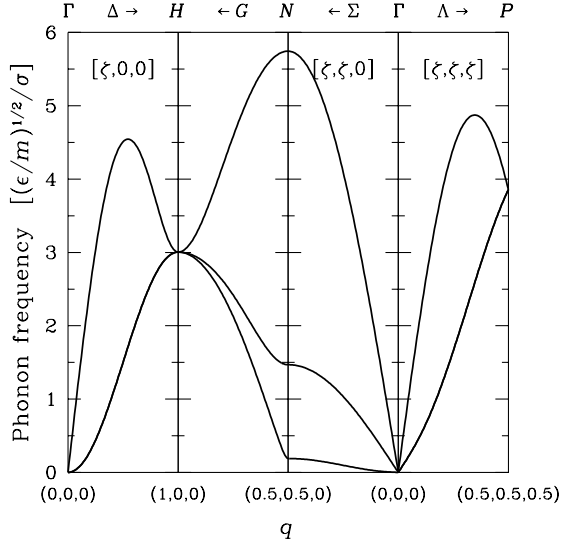}
\caption{Yoshida-Kamakura potential (\ref{3-12}) with $a=3.3$:
phonon branches of the BCC crystal for $\rho=0.6607$ along a number of
high-symmetry lines in {\bf q} space. Along the $\Gamma$N path, one of
the branches is seen to soften at the $\Gamma$ point due to the vanishing
of $c_{44}$.}
\label{fig3}
\end{figure}

Upon varying the value of $a$ in Eq.\,(\ref{3-12}), one can follow the
evolution of the YK phase diagram through the elastic
criterion of melting~\cite{Prestipino}. For
large $a$ values, the inverse-power-fluid limit is recovered; for $a\simeq 7$,
there appears a region of bcc stability between the low- and high-density fcc
solids; on decreasing $a$ more and more, the stable-bcc region gradually
shrinks until, for $a\approx 4$, a gap opens between the bcc and high-density
fcc regions, signalling the stabilization for intermediate pressures of one
or more crystals of symmetry other than cubic. The opening of the gap is
preceded by the onset of reentrant melting, which first occurs
for $a\approx 5$.

Another instance of core-softened repulsion is provided by the modified
inverse-power (MIP) potential studied in Ref.\,~\cite{Malescio}.
The following one-parameter family of potentials is being considered:
\be
\phi_{\rm MIP}(r)=\epsilon\left(\frac{\sigma}{r}\right)^{n(r)}
\,\,\,\,{\rm with}\,\,n(r)=12
\left\{1-a\exp\left[-5\left(1-\frac{r}{\sigma}\right)^2\right]\right\}\,,
\label{3-13}
\ee
where $0<a<1$ is a softness parameter, {\it i.e.}, a number fixing the extent
to which the inverse-power exponent deviates from 12 in the close
neighbourhood of $\sigma$. Upon increasing $a$, the potential core softens
more and more, with the effect of destabilizing both the fcc and the bcc
order for intermediate densities. This is accompanied by reentrant melting
and by the appearance of one or more
low-coordinated crystal phases in the pressure gap left open by bcc and fcc.
In the left panel of Fig.\,4, I report the phase diagram of the MIP fluid for
$a=0.8$ as obtained from Monte Carlo simulation through the heat-until-it-melts
method~\cite{Malescio};
the same melting lines but derived from the elastic criterion are plotted
in the right panel of Fig.\,4. Again, we see more than one correspondence
between the present melting criterion and the simulation results.

\begin{figure}
\includegraphics[width=8cm]{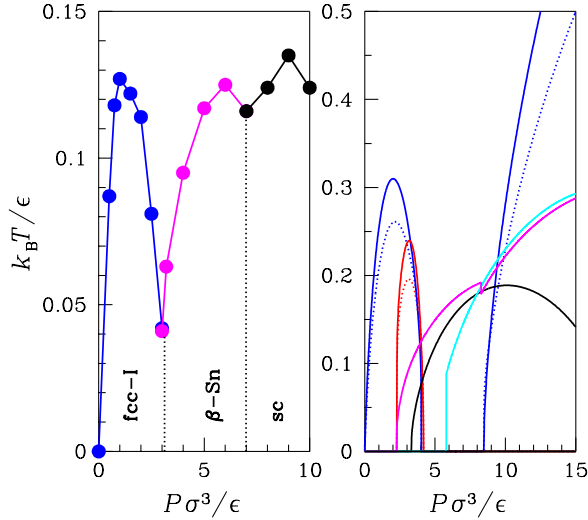}
\caption{(Color online). Modified inverse-power potential with $a=0.8$.
Left: Numerically-computed phase diagram (reprinted from
Ref.\,\cite{Malescio}; the dots are melting points as obtained by the
heat-until-it-melts method while the vertical dotted lines are putative
solid-solid boundaries as extrapolated from exact total-energy calculations
at $T=0$); right: same phase diagram as predicted through
the elastic criterion of melting (blue, fcc; red, bcc; black, sc; cyan, sh;
magenta, $\beta$-Sn; the blue and red dotted lines are the melting curves
for the fcc and bcc crystals, respectively, as drawn from the harmonic
approximation).}
\label{fig4}
\end{figure}

However, there are also instances (arguably not so common) where the elastic
criterion fails badly. This occurs when a crystal that is predicted by linear
elasticity to be unstable at $T=0$ is in fact stabilized in a range of
temperatures, somewhat counterintuitively, by virtue of anharmonic effects.
In a case of these, anharmonicity manages to make a crystal phase rigid to
small deformations in spite of the violation of the stability conditions of
elasticity. I found one case of these for the MIP potential.
The $T=0$ calculation of the bcc elastic moduli for $a=0.6$ predicts a gap of
stability in the density range from $\rho=0.910$ ($P\simeq 5.761$) to
$\rho=1.066$ ($P\simeq 8.168$), whereas for e.g. $\rho=1$ ($P\simeq 7.021$ at
$T=0$) Monte Carlo simulation clearly indicates that the bcc solid is stable
up to $T\simeq 0.105$~\cite{Malescio} (all quantities in reduced units).
A numerical calculation of the elastic constants for $\rho=1$ at very low
temperature ($T=0.001$) with the method of Ref.\,\cite{Farago} indeed reveals
large deviations from the $T=0$ values, which is not the case for e.g.
$\rho=0.7$ ($P\simeq 2.846$ at $T=0$),
where the agreement with linear elasticity is much better (see Table 1).
What is happening then? The similar situation with Calcium sc phase
provides a clue~\cite{Errea}: strong enough anharmonic terms in the Hamiltonian
(classical or quantum) may succeed to convert imaginary phonon frequencies into
real ones, thus allowing the alleged unstable solid to become mechanically
(and thermodynamically) stable.

\begin{table}
\caption{MIP potential for $a=0.6$, elastic constants of the BCC crystal at
the reduced densities $\rho=0.7$ and $\rho=1$. The exact $T=0$ values derived
from Eqs.\,(\ref{2-6}) are compared with their MC estimates at $T=0.001$
(for samples of $N=686$ particles and equilibrium trajectories of as many
as $2\times 10^5$ MC moves per particle). While the BCC crystal would
be mechanically unstable at $\rho=1$ according to elasticity theory, it is
actually found perfectly rigid to thermal fluctuations in numerical simulation
owing to the stabilizing effect of the anharmonicities in the potential.}
\bigskip
\begin{tabular}{cccccccc} \\
\hline
& & $\rho=0.7$ & & & & $\rho=1$ \\
\hline
& $c_{11}$ & $c_{12}$ & $c_{44}$ & & $c_{11}$ & $c_{12}$ & $c_{44}$ \\
\hline
\hline
$T=0$ & 10.06156 & 8.64219 & 2.94964 & & 18.92754 & 13.80805 & $-0.41631$ \\
MC & 10.046(1) & 8.632(1) & 2.937(1) & & 18.1(3) & 13.89(3) & 0.955(4) \\
\hline
\end {tabular}
\end{table}

%
%
\section{The melting curve of the Gaussian-core model}
\setcounter{equation}{0}
\renewcommand{\theequation}{4.\arabic{equation}}
The Gaussian-core model (GCM) fluid ({\it i.e.}, classical point particles
interacting through a repulsive Gaussian potential in three dimensions)
gives the opportunity to compare the relative efficacy of various empirical
melting rules, all rooted in the use of the Lindemann criterion.
In particular, we shall figure out the merits and
drawbacks of the self-consistent harmonic approximation (SCHA)~\cite{scha},
which for many years represented a popular theoretical alternative to exact
free-energy calculations.

Besides a fluid phase, the GCM shows two distinct, fcc and bcc solid
phases~\cite{Stillinger}. At $T=0$, the fcc solid transforms to bcc for
$P=0.05529$. At higher pressures and for $T>0$, the bcc solid undergoes
reentrant melting: $T_m(P)$ is an increasing function
for $P\lesssim 0.136$ while being decreasing otherwise, further vanishing in
the limit of infinite pressure. The fcc and bcc melting lines as predicted by
the elastic criterion are reported in Fig.\,5, together with those obtained
from the harmonic approximation.
In the same picture, the outcome of a variational treatment~\cite{Lang} and the
numerically-computed coexistence lines~\cite{Prestipino2} are also plotted for
comparison. Clearly, the simple elastic criterion is able to account for the
main characteristics of GCM melting, though the fcc and bcc melting
temperatures are again found to be about twice larger than the actual values
and the threshold where the fcc solid is
overcome in stability by the bcc phase remains vague, much overestimated by the
putative fcc reentrant-melting line. Quantitatively speaking, the harmonic
approximation and, especially, the variational theory provide more valid
alternatives to free-energy calculations.

\begin{figure}
\includegraphics[width=8cm]{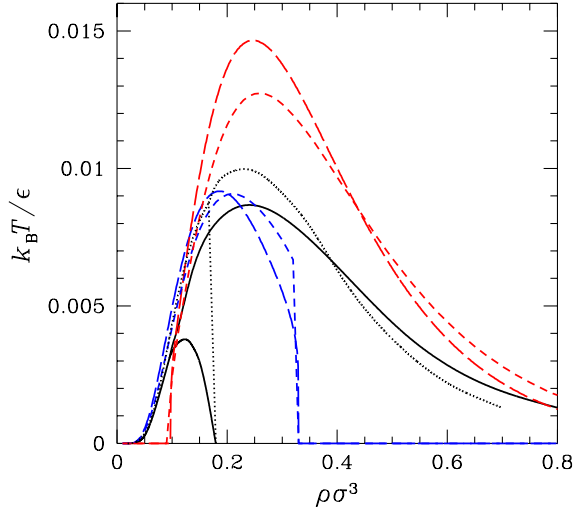}
\caption{(Color online). The Gaussian-core model phase diagram as determined
through various methods: exact free-energy calculations (solid black lines);
variational method (dotted black lines); elastic criterion of melting
(long-dashed blue and red lines -- blue, fcc; red, bcc); harmonic
approximation (dashed blue and red lines).}
\label{fig5}
\end{figure}

The SCHA is a theory for the thermal attenuation of phonon energies that aims
at introducing elements of anharmonicity in an otherwise harmonic set-up.
It provides an internal, self-consistent condition for its own validity which
had sometimes been interpreted as an indication of the maximum temperature at
which the crystal can be superheated. When used in combination with the
Lindemann rule, the SCHA provides an independent melting criterion.
Before illustrating the specific prediction for the GCM,
I present a brief introduction to the SCHA. 

The formal justification of the SCHA lies in the use of the variational method
of statistical mechanics. The strategy is focussed on determining the
``optimal'' harmonic approximation to the real Hamiltonian at the given
temperature $T$, which is generally not its harmonic part. The crucial
assumption is that of an integrable pair potential $\phi({\bf r})$, endowed
with a Fourier transform $\widetilde{\phi}({\bf q})$. This automatically
excludes hard-core potentials, for which the SCHA theory cannot be formulated.
The average of the system potential energy over a reference harmonic system
$U_{\rm harm}$, having the same potential-energy minimum as the system of
interest but {\em different} phonon frequencies $\omega_s({\bf k})$ and
normal-mode amplitudes $\boldsymbol{\epsilon}_s({\bf k})$, is
\be
\left<U\right>_{\rm harm}=\frac{v_0}{2}\sum_{i,j}^\prime
\int\frac{{\rm d}^3q}{(2\pi)^3}\widetilde{\phi}({\bf q})e^{i{\bf q}\cdot
({\bf R}_i-{\bf R}_j)}e^{-\frac{1}{2}\left<\left({\bf q}\cdot
({\bf u}_i-{\bf u}_j)\right)^2\right>_{\rm harm}}\,,
\label{4-1}
\ee
where the prime over the sum means $i\neq j$ and
\be
\frac{1}{2}\left<\left({\bf q}\cdot
({\bf u}_i-{\bf u}_j)\right)^2\right>_{\rm harm}=\frac{k_BT}{N}
\sum_{{\bf k},s}\left({\bf q}\cdot\boldsymbol{\epsilon}_s({\bf k})\right)^2
\frac{1-e^{i{\bf k}\cdot({\bf R}_i-{\bf R}_j)}}{m\omega_s^2({\bf k})}\equiv
D({\bf q},\{{\bf R}\})\,.
\label{4-2}
\ee
The best approximation to the Helmholtz free energy of the system within all
conceivable harmonic interactions is given by the minimum of the
Gibbs-Bogoliubov functional,
\be
\widetilde{F}[H_{\rm harm}]\equiv F_{\rm harm}+
\left<H-H_{\rm harm}\right>_{\rm harm}=
F_{\rm harm}+\left<U\right>_{\rm harm}-U_0-\frac{3}{2}Nk_BT\,,
\label{4-3}
\ee
where the Helmholtz free energy of the reference system reads
\be
F_{\rm harm}=U_0+3Nk_BT\ln\left(\frac{\Lambda}{v_0^{1/3}}\right)+\frac{k_BT}{2}
\sum_{{\bf k},s}\ln\left(\frac{m\omega_s^2({\bf k})v_0}{\pi k_BT}\right)
\label{4-4}
\ee
with $\Lambda$ the thermal wavelength.
Using the frequencies $\omega_s({\bf k})$ as variational parameters,
they are eventually obtained as the solutions of the SCHA equations
\be
m\omega_s^2({\bf k})=v_0\sum_{j\ne 1}\left(e^{-i{\bf k}\cdot
{\bf R}_j}-1\right)\int\frac{{\rm d}^3q}{(2\pi)^3}\left({\bf q}\cdot
\boldsymbol{\epsilon}_s({\bf k})\right)^2\widetilde{\phi}({\bf q})
e^{-i{\bf q}\cdot{\bf R}_j}e^{-D({\bf q},\{{\bf R}\})}\,.
\label{4-5}
\ee
In practice, the target temperature $T$ is reached in steps, where at every
step of the calculation the equations (\ref{4-5}) are solved iteratively
until the left-hand side equates to a certain degree of precision the
right-hand side.
At low temperature, a good starting point of the iteration are the system
own frequencies. Observe that, thanks to symmetry considerations, a (congruous)
number of ${\bf k}$ vectors in a small fraction of the 1BZ will suffice for
the calculation of a sum like that in $D$ (e.g. just 1/48 of the full 1BZ
for the FCC lattice)~\cite{Wallace}. Once $D$ is obtained, the matrix
\be
Z_{\alpha\beta}({\bf k})=v_0\sum_{j\neq 1}\left(\cos({\bf k}\cdot
{\bf R}_j)-1\right)\int\frac{{\rm d}^3q}{(2\pi)^3}\,q_\alpha q_\beta
\widetilde{\phi}({\bf q})\cos({\bf q}\cdot{\bf R}_j)e^{-D}
\label{4-6}
\ee
is diagonalized in order to extract its eigenvalues $m\omega_s^2({\bf k})$ and
eigenvectors $\boldsymbol{\epsilon}_s({\bf k})$, and this completes a single
iteration step.

\begin{figure}
\includegraphics[width=8cm]{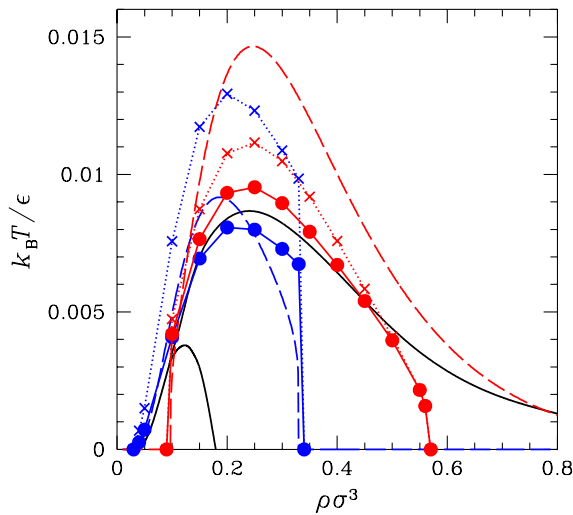}
\caption{(Color online). Gaussian-core model phase diagram: the outcome of
the SCHA (blue and red dots, joined by solid straight lines -- blue, fcc;
red, bcc) is compared with that of the elastic criterion of melting
(long-dashed blue and red lines). The SCHA instability temperatures for the
fcc and bcc solids are also plotted as blue and red crosses, respectively,
joined by dotted straight lines. Finally, the solid black lines mark the
exact coexistence loci.}
\label{fig6}
\end{figure}

The main limitation in the use of the SCHA method is computational, due to
the necessity of solving numerically a large number of times the integral in
(\ref{4-6}) to a high degree of precision. In the GCM case this integral can
be computed analytically and this enormously speeds up the whole procedure.
Even in this favourable situation, computing a single melting point by
the SCHA method takes a time typically three orders of magnitude larger
than if we apply the elastic criterion, which performs the calculation
in a few hundredths of a second on a fast PC.
In general, for a given density $\rho$ the self-consistent calculation of the
frequencies $\omega_s({\bf k})$ and the respective MSD can be accomplished
only up to a certain temperature $T_i(\rho)$, which is called the instability
temperature. Beyond this temperature, no self-consistent solution of the
Eqs.\,(\ref{4-5}) is found. Moreover, depending on $\rho$ the ratio of the MSD
at $T_i$ to $a_{NN}$ may even exceed $L_m$ (in this event, I assume $T_m=T_i$).
The SCHA results for the GCM are reported in Fig.\,6. Compared to the outcome
of the elastic criterion, the SCHA estimate of the GCM melting temperature is
better for all low to intermediate densities; the SCHA is instead unable to
reproduce the large-density tail of $T_m(P)$ since beyond a density of 0.57
I find no self-consistent solution of the Eqs.\,(\ref{4-5}) and the
relative $T_m$ hence drops to zero.
It is worth adding a final remark about the SCHA instability thresholds for
the GCM model. As we see from Fig.\,6, the $T_i$ for fcc is higher than it
is for bcc, in sharp contrast with the sequence of melting thresholds.
In fact, the SCHA instability at, say, $\rho\sigma^3=0.2$ occurs, for both
phases, where the root mean square displacement (rmsd) for the reference
system is roughly a fraction 0.23 of the nominal NN distance ($r_{NN}$).
But the rate of growth with temperature of rmsd$/r_{NN}$ is slightly larger
for bcc, with a pronounced acceleration above a level of about 0.17 for bcc
and 0.20 for fcc; hence, the rmsd$/r_{NN}$ of the bcc crystal reaches the
values 0.18 (melting) and 0.23 (instability) both within the range comprised
between the fcc melting and instability temperatures.

%
%
\section{Conclusions}
Through the use of representative model potentials, I managed to show that
a simple melting criterion based on the Lindemann rule and a description of
the solid as an elastic medium is able to capture, with negligible
computational effort, the overall characteristics of the system melting line.
In more quantitative terms, the criterion overestimates the melting
temperature by roughly a factor of two for fcc and bcc solids, independently
of the pressure value. For other crystals, the prediction of the criterion is
less reliable, mainly due to the uncertainty on the value of the Lindemann
parameter and its actual pressure dependence. In fact, the value of the
elastic criterion of melting is more of a heuristic kind, {\it i.e.}, of
guidance for fastly detecting the existence of anomalies in the melting line,
as in case of reentrant-fluid behaviour in a system where the softness of the
particle core can be made to vary by tuning an appropriate parameter in the
potential. The accuracy obtained by the elastic criterion in predicting
the overall appearance of the phase diagram can be comparable to that of more
sophisticated (two-phase) theories of fluid-solid coexistence, as I showed
for one instance of core-softened interaction.
For a Gaussian repulsive core, I compared the outcome of the elastic criterion
with the harmonic approximation, as well as with the more effective but also
more numerically demanding self-consistent harmonic approximation. Though
moving upward in the hierarchy of theories generally improves the estimate
of the melting temperature for all pressures, the gain in accuracy is only
marginal and, more important, the topology of the melting line stays
unaltered. Hence, at least for the Gaussian potential, a description in
terms of zero-temperature elastic constants is by far sufficient to anticipate
the essential features of the melting behaviour and no better theory is
strictly necessary.

%
%
\appendix
\section{A statistical theory of the fluid-solid transition}
\setcounter{equation}{0}
\renewcommand{\theequation}{A.\arabic{equation}}
In this Appendix, a theory of fluid-solid coexistence is formulated for
purely repulsive potentials, where the fluid phase is described through
the variational approach by Mansoori and Canfield~\cite{MC} while the
statistical properties of each solid phase are modelled through a cell theory.

Assuming the hard-sphere (HS) fluid as reference, it derives from the
Gibbs-Bogoliubov inequality that the exact Helmholtz free energy per
particle $f$ of a system with potential $\phi(r)$ is bounded from above by
\be
f^*(T,v;\sigma_{\rm HS})=f_{\rm HS}+\frac{\rho}{2}\int{\rm d}^3r\,
g_{\rm HS}(r)w(r)\,\,\,\,{\rm with}\,\,w(r)=\phi(r)-\phi_{\rm HS}(r)\,,
\label{a-1}
\ee
where $\rho=1/v$ is the number density and $g_{\rm HS}(r)$ is the HS radial
distribution function (RDF). In Eq.\,(\ref{a-1}), the HS particle diameter
$\sigma_{\rm HS}$ is left unspecified; the best approximant to $f$ is obtained
by minimizing (\ref{a-1}) with respect to $\sigma_{\rm HS}$. Although the HS
equation of state is not known exactly, a good approximation is the
Carnahan-Starling form~\cite{Hansen} from which the HS free energy follows as
\be
f_{\rm HS}=k_BT\left[\ln(\rho\Lambda^3)-1\right]+
k_BT\frac{\eta(4-3\eta)}{(1-\eta)^2}
\label{a-2}
\ee
with $\eta=(\pi/6)\rho\sigma_{\rm HS}^3$. To obtain an estimate of the HS RDF,
one resorts to the Percus-Yevick approximation~\cite{Hansen} since then the
direct correlation function $c_{\rm HS}(r)=c_0(r/\sigma_{\rm HS};\eta)$
is known in a closed form:
\ba
c_0(x)&=&\left\{
\begin{array}{ll}
-\lambda_0-\lambda_1x-\lambda_3x^3 & ,\,\,x<1 \\
0 & ,\,\,x\ge 1
\end{array}
\right.
\nonumber \\
&& {\rm with}\,\,\lambda_0=\frac{(1+2\eta)^2}{(1-\eta)^4}\,,
\,\,\lambda_1=-6\eta\frac{(1+\eta/2)^2}{(1-\eta)^4}\,,
\,\,\lambda_3=\eta\frac{\lambda_0}{2}\,.
\label{a-3}
\ea
The Ornstein-Zernike relation then yields
$g_{\rm HS}(r)=g_0(r/\sigma_{\rm HS};\eta)$ with
\be
g_0(x)=1+\frac{2}{\pi}\int_0^\infty{\rm d}k\,k^2\frac{\sin(kx)}{kx}
\frac{\widetilde{c}_0(k)}{1-24\eta\widetilde{c}_0(k)}\,\,\,\,{\rm and}\,\,
\widetilde{c}_0(k)=\int_0^1{\rm d}x\,x^2\frac{\sin(kx)}{kx}c_0(x)\,.
\label{a-4}
\ee
The variational free energy (\ref{a-1}) can then be written as
\ba
f^*&=&k_BT\left[\ln(\rho\Lambda^3)-1\right]+
k_BT\frac{\eta(4-3\eta)}{(1-\eta)^2}+12\eta\int_1^\infty{\rm d}x\,x^2
g_0(x;\eta)\phi(x\sigma_{\rm HS})
\nonumber \\
&\equiv&3k_BT\ln\frac{\Lambda}{\sigma}+k_BT\left[\ln(\rho\sigma^3)-1\right]+
\Delta f^*\,,
\label{a-5}
\ea
where $\sigma$ is an arbitrary length unit.
Called $\bar{\sigma}_{\rm HS}(T,v)$ the optimal $\sigma_{\rm HS}$ value
and observing that $\Delta f^*$ depends on $v$ only through $\eta$
({\it i.e.}, $\Delta f^*(T,v;\sigma_{\rm HS})=
\varphi(T,\eta(v,\sigma_{\rm HS});\sigma_{\rm HS})$), the fluid chemical
potential can be approximated as $\mu=\bar{f}+\bar{P}v$, where
$\bar{f}=f^*(T,v;\bar{\sigma}_{\rm HS}(T,v))$ and
$\bar{P}=-\partial\bar{f}/\partial v$.
In order to calculate $\bar{P}$, one considers that
\be
\left.\frac{\partial\Delta f^*}{\partial\sigma_{\rm HS}}\right|_{T,v}=0
\,\,\,\,{\rm whence}\,\,
\left.\frac{\partial\varphi}{\partial\eta}\right|_{T,\sigma_{\rm HS}}=
-\frac{\sigma_{\rm HS}}{3\eta}\left.
\frac{\partial\varphi}{\partial\sigma_{\rm HS}}\right|_{T,\eta}\,.
\label{a-6}
\ee
As a result,
\ba
\bar{P}v&\equiv&-v\left.\frac{\partial\bar{f}}{\partial v}\right|_T=
k_BT-v\left.\frac{\partial\Delta\bar{f}}{\partial v}\right|_T=
k_BT-\frac{\sigma_{\rm HS}}{3}\left.
\frac{\partial\varphi}{\partial\sigma_{\rm HS}}\right|_{T,\eta}
\nonumber \\
&=&k_BT-\frac{2\pi}{3}\frac{\sigma_{\rm HS}^4}{v}\int_1^\infty{\rm d}x\,
x^3g_0(x;\eta)\phi'(x\sigma_{\rm HS})\,.
\label{a-7}
\ea
This completes the derivation of an approximate expression of the fluid
chemical potential to be compared with the chemical potential of
the solid phase.

\begin{figure}
\includegraphics[width=8cm]{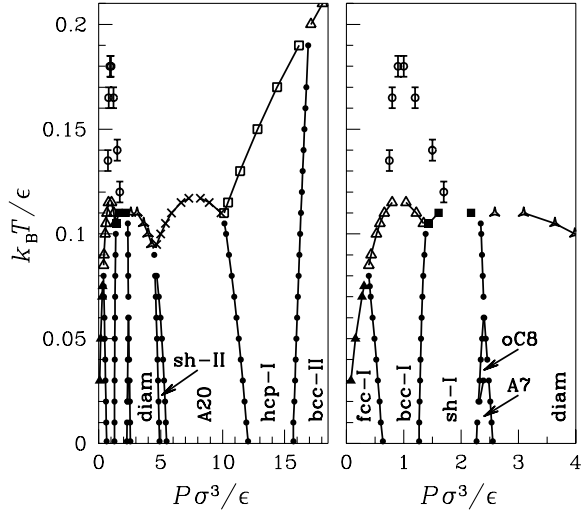}
\caption{Left: phase diagram of the YK potential with $a=2.1$
according to the theory detailed in the Appendix (for the meaning
of Roman numerals, see Fig.\,2 caption). Right: zoom on the
low-pressure region. Solid-solid coexistence points are depicted as
small dots, whereas triangles, squares, tripods, and crosses are
solid-fluid coexistence points. See Ref.\,\cite{Saija2} for a
comparison with the prediction from Monte Carlo simulation. The
open dots with error bars give the location of number-density maxima
within the Mansoori-Canfield description of the fluid phase.}
\label{fig7}
\end{figure}

As far as the solid sector of the phase diagram is considered, I first
determine the stable phases at $T=0$ through a series of total-energy
calculations for a large number of candidate crystal structures (see
Ref.\,\cite{Prestipino3} for more details). To obtain a rough estimate
of the crystal chemical potential at $T>0$, I use
the simple Lennard-Jones-Devonshire cell theory~\cite{celltheory}.
In this theory, a crystal partition function of effectively independent
particles is written down where any given particle, which can be
found anywhere in its own Wigner-Seitz cell (WSC), is acted
upon by the force exerted by the other $N-1$ particles, placed at
equilibrium lattice positions. In practice, the canonical partition function
of a crystal is approximated as
\be
Z=\frac{1}{\Lambda^{3N}}\int_{{\rm WSC}_1}{\rm d}^3r_1\cdots\int_{{\rm WSC}_N}
{\rm d}^3r_N\,\exp\left\{-\sum_i\widetilde{\phi}({\bf r}_i)/(k_BT)\right\}\,,
\label{a-8}
\ee
where
\be
\widetilde{\phi}({\bf r})=\frac{1}{2}\sum_{j\neq 1}\phi(|{\bf R}_1-{\bf R}_j|)+
\sum_{j\neq 1}\left[\phi(|{\bf R}_1+{\bf r}-{\bf R}_j|)-
\phi(|{\bf R}_1-{\bf R}_j|)\right]\,.
\label{a-9}
\ee
Taking
\be
\Phi({\bf r})=\sum_{j\neq 1}\phi(|{\bf R}_1+{\bf r}-{\bf R}_j|)
\,\,\,\,{\rm and}\,\,
\Psi({\bf r})=-\frac{1}{3}\sum_{j\neq 1}|{\bf R}_1+{\bf r}-{\bf R}_j|
\phi'(|{\bf R}_1+{\bf r}-{\bf R}_j|)\,,
\label{a-10}
\ee
a direct calculation offers
\be
\frac{F}{N}=-k_BT\ln\frac{v_f}{\Lambda^3}+\frac{1}{2}\Phi(0)
\,\,\,\,{\rm with}\,\,
v_f=\int_{\rm WSC}{\rm d}^3r\,\exp\{-\beta\left[\Phi({\bf r})-\Phi(0)\right]\}
\label{a-11}
\ee
and
\ba
\mu=\frac{F}{N}+Pv&=&3k_BT\ln\frac{\Lambda}{\sigma}-
k_BT\left(\ln\frac{v_f}{\sigma^3}-1\right)+\frac{1}{2}
\left[\Phi(0)+\Psi(0)\right]
\nonumber \\
&+&\frac{\int_{\rm WSC}{\rm d}^3r\left[\Psi({\bf r})-\Psi(0)\right]
\exp\left\{-\beta[\Phi({\bf r})-\Phi(0)]\right\}}
{\int_{\rm WSC}{\rm d}^3r\exp\left\{-\beta[\Phi({\bf r})-\Phi(0)]\right\}}\,.
\label{a-12}
\ea

Fig.\,7 (left panel) shows the phase diagram of the Yoshida-Kamakura
potential (\ref{3-12}) for $a=2.1$ as mapped out in the way just explained.
A zoom on the
low-pressure region of the phase diagram is presented in the right panel of
Fig.\,7. Compared to the exact phase diagram of Ref.~\cite{Saija2}, we see that
the theory correctly accounts for the succession and extent of solid phases
(with the unique omission of the cI16 solid), though still overestimating the
values of the melting temperature by approximately 100\% for all pressures.
In the same picture, I also plotted the line encompassing the region of density
anomaly as computed within the Mansoori-Canfield theory. The shape of this
line compares well with that of the same line as obtained from simulation.


\begin{thebibliography}{99}
\bibitem{nota} A short presentation of the method has been given recently
in Ref.\,\cite{Prestipino}.
\end{thebibliography}
\end{document}